\documentclass[twocolumn,prb,showpacs]{revtex4}
\usepackage{amsmath, graphicx}
\newcommand{\vect}[1]{\mathbf{#1}}
\begin{document}
\author{Jonathan Keeling}
\affiliation{Cavendish Laboratory, University of Cambridge,
  J.~J.~Thomson Ave., Cambridge CB3 0HE, UK}
\email{jmjk2@cam.ac.uk}
\title{Polarized polariton condensates and coupled XY models}
\begin{abstract}
  Microcavity polaritons, which at low temperatures can condense to a
  macroscopic coherent state, possess a polarization degree of
  freedom.  This article discusses the phase diagram of such a system,
  showing the boundaries between differently polarized condensates.
  The Bogoliubov approximation is shown to have problems in describing
  the transition between differently polarized phases; the
  Hartree-Fock-Popov approximation performs better, and compares well to
  exact results that can be used in the limit where the left- and
  right-circular polarization states decouple.  The effect on the
  phase boundary of various symmetry breaking terms present in real
  microcavities are also considered.
\end{abstract}
\maketitle

\section{Introduction}
\label{sec:introduction}

The recent experimental progress in attaining spontaneous coherence in
a thermalized degenerate gas of microcavity
polaritons~\cite{kasprzak06:nature,snoke06:condensation,kasprzak07,marchetti07,love08,lagoudakis08,utsunomiya08}
has extended the range of systems in which quantum condensation may be
studied.
As well as sharing many features with previous examples of quantum
condensates (such as superfluid Helium, cold atoms and
superconductors) polariton condensates possess naturally a number of
distinguishing features (see e.g.
Refs.~\onlinecite{kavokin07,keeling_review07}, and Refs. therein).

This article considers the combined effect of two such features of
condensed polaritons: their polarization degree of freedom and
confinement to two dimensions.
Recent works~\cite{shelykh06,rubo06,rubo07} have considered some
effects arising from the polarization degree of freedom; including a
tentative phase diagram~\cite{rubo06} derived from a zero-temperature
mean-field theory, this phase diagram implied a transition temperature
which vanished at a critical magnetic field.
This article has two aims: first to examine more carefully the phase
diagram (critical temperature vs magnetic field) of a polarized
polariton condensate, this is found to be rather different to that in
Ref.~\onlinecite{rubo06}; second to discuss how the phase diagram
would be affected by terms breaking polarization rotation symmetry
that are expected in real microcavities.

In order to discuss the effects of polarization on the polariton phase
diagram in the most transparent way, this article makes various
simplifications: it considers an infinite two-dimensional polariton
system, in thermal equilibrium.
In addition, the model described in Refs.~\onlinecite{rubo06,rubo07}
is used, which is applicable in the limit of very low densities and
temperatures, where only the low energy part of the lower polariton
dispersion --- with a quadratic dispersion --- is thermally populated.
At higher densities, one must take account of the non-quadratic
dispersion of lower polaritons, and the possibility of other
excitations depleting the condensate~\cite{marchetti07}.
These considerations will change the dependence of critical
temperature on density, but the nature of the possible polarized
phases and topology of the phase diagram as a function of magnetic
field should not significantly change.

Current experiments remain some distance from this limit of low
temperature, low density, infinite, clean equilibrium systems: The
experimental densities are at the point where other excitations start
to become relevant~\cite{marchetti07}, the polariton clouds are
relatively small and their profiles strongly affected by photonic
disorder~\cite{kasprzak07,love08,lagoudakis08} or trapping
potentials~\cite{snoke06:condensation}, and pumping and decay have
noticeable effects\cite{marchetti07,love08,lagoudakis08}.
It does however remain a useful exercise to understand how spin
degrees of freedom modify the phase diagram in the ideal system first,
before including these various extra complications, both to understand
the nature of the possible phases and transitions, and also to have a
basis from which one can investigate the differences introduced by
pumping, decay, disorder and finite sizes.
In addition, improvements in fabrication of microcavity samples can be
expected to reduce the effect of disorder, and increase the lifetime
of polaritons, opening the possibility that future generations of
experiments might come closer to the idealized model discussed here.
A review of how some of these effects of disorder, pumping, decay etc
affect coherence in current microcavity polariton experiments can be
found for example in Ref.~\onlinecite{keeling_review07}.

\section{The Model}
\label{sec:model}


%
The polarization of the polariton can be written as a two-component complex
spinor $\vec{\psi}$; 
\begin{equation}
  \label{eq:param}
  \vec{\psi}
  =
  \left(
    \begin{array}{cc}
      \psi_x \\ \psi_y
    \end{array}
  \right)
  =
  \frac{l}{\sqrt{2}} \left(
    \begin{array}{cc}
      1 \\ i
    \end{array}
  \right)
  +
  \frac{r}{\sqrt{2}} \left(
    \begin{array}{rr}
      1 \\ -i
    \end{array}
  \right)
\end{equation}
Here $l$ and $r$ are complex coefficients, describing the state in the
basis of left- and right-circular polarizations. The model
of~\citet{rubo06} in this basis is:
\begin{multline}
  \label{eq:hamil-fg}
  H-\mu N = \frac{\hbar^2 |\nabla l|^2}{2m} + \frac{\hbar^2 |\nabla r|^2}{2m}
  - (\mu + \Omega) |l|^2 - (\mu - \Omega) |r|^2 
  \\
  + \frac{1}{2}
  \left[
    U_0 \left( |l|^4 + |r|^4 \right)  + (U_0 - 2 U_1) 2|l|^2 |r|^2 
  \right]
\end{multline}
The term $\Omega$ describes a magnetic field that favors either left-
or right-circular polarization.
From this point onwards $\hbar=1$.
In 2D the form of the phase diagram is controlled by two dimensionless
parameters; $m U_0$ and $(U_0 - 2 U_1)/U_0$.
For $m U_0$, estimates including effects of excitonic
disorder~\cite{marchetti07} give $U_0 \simeq 3 \mu \text{eV}
(\mu\text{m})^2$, and the polariton mass $1/m = 7600 \mu \text{eV}
(\mu\text{m})^2$, lead to $m U_0 \simeq 4 \times 10^{-4}$.
Due to the tendency toward biexciton formation, opposite polarizations
of polaritons attract, hence $0 > U_0 - 2U_1 > -U_0$; at $\Omega=0$
this implies $|l|=|r|$ describing linear polarization of
light~\cite{rubo06}.
For the typical values relevant for microcavity polaritons $|U_0 -
2U_1| \ll U_0$; i.e.\ the interaction between left- and
right-circularly polarized light is relatively small~\cite{renucci05};
a typical estimate is $(U_0- 2 U_1) = - 0.1 U_0$.
The model of Eq.~(\ref{eq:hamil-fg}) has been studied in the context
of atomic condensation, e.g. Ref.~\onlinecite{battye02}, where topological
defects and phase separation were investigated; as discussed there,
phase separation requires $U_1 < 0$, so is not considered here.

\section{Calculating critical temperature}
\label{sec:calc-crit-temp}

In an infinite two-dimensional system, for a single component Bose
gas, it is well known that the phase transition is a
Berezhinskii-Kosterlitz-Thouless (BKT) transition.
Above the BKT transition vortices proliferate, leading to
exponentially decaying correlations; below there is a non-zero
superfluid density, and a quasi-condensate density, but at long
distances phase fluctuations lead to power law decay of correlations.
To see how this is modified in a spinor Bose gas, one needs to
consider the elementary vortices of the model in
Eq.~(\ref{eq:hamil-fg}).
As discussed by~\citet{rubo07}, these elementary vortices are separate
vortices of left- and right-circularly polarized light.
Thus, to a first approximation, the critical temperature occurs when
the phase stiffness of $l$ or of $r$ becomes small enough that
vortices of $l$ or $r$ will proliferate.
Note however that although the elementary vortices are separate
vortices of $l$ and $r$, as long as $U_0-2U_1\ne 0$ then the long
wavelength phase fluctuations are mixtures~\cite{rubo06,shelykh06} of
$l$ and $r$.

When $l$ and $r$ vortices are independent, the critical temperature at
which vortices proliferate is given by $\rho_s^{l,r} = (2/\pi) m k_B
T$ where $\rho_s^{l,r}$ are the phase stiffness of each species.
Section~\ref{sec:symm-break-effects} will show that in the absence of
symmetry breaking terms beyond the model in Eq.~(\ref{eq:hamil-fg}),
it is appropriate to treat these transitions independently, and will
discuss how symmetry breaking terms modify this argument.
This article will present the phase diagram of critical temperature as
a function of magnetic field, working at a fixed total density of
polaritons.
With such an approach, to calculate the phase boundary one will
require expressions for the total density
$\rho_{\text{total}}(\mu,\Omega,T)$ of the coupled system, and for the
superfluid densities for each separate polarization $\rho_{s}^{l,r}$;
the following sections will discuss calculating these quantities.

\subsection{Decoupled case, $U_0=2U_1$}
\label{sec:deco-case-u_0=2}

The interaction between left and right polarizations is relatively
weak, and so it is worth first considering the special case $U_0 =
2U_1$, for which the polarizations decouple.
This case is simple for two reasons:
Firstly, it is clear that there are two completely independent phase
transitions associated with BKT transitions for the $l$ and $r$
polarizations.
Secondly, one may reuse the equation of state $\rho_{1}(\mu,T)$ and
critical chemical potential $\mu_{c1}$ for a one-component 2D Bose
gas, which allows a comparison between the exact equation of state
and various perturbative approximations.

\begin{figure}[htpb]
  \centering
  \includegraphics[width=0.9\columnwidth]{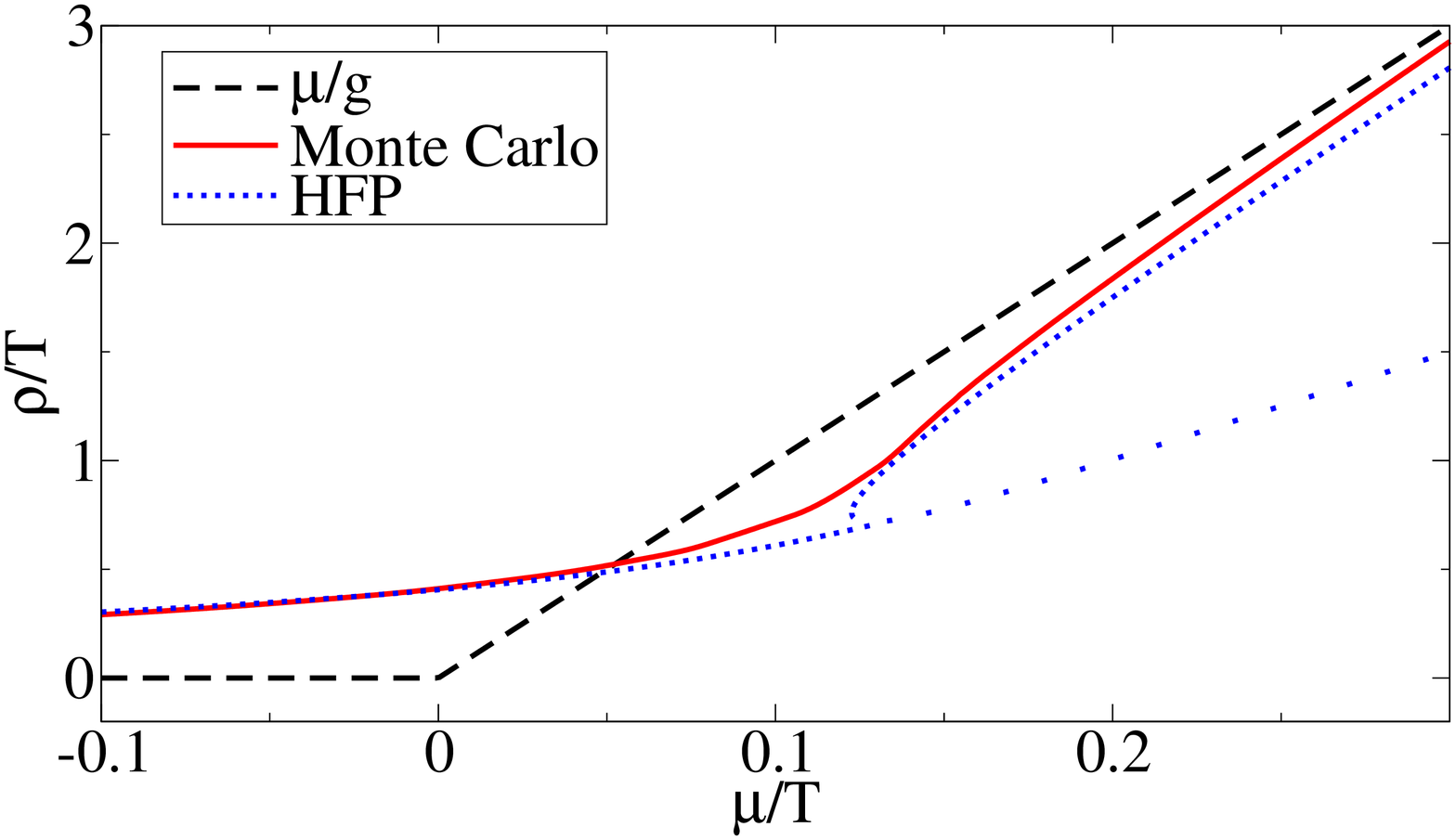}
  \caption{(Color online) Equation of state, $\rho(\mu,T) = T
    f(\mu,T)$ for a 2D single component Bose gas, as given by HFP
    approximation and Monte Carlo results of
    Ref.~\onlinecite{prokofev02}. The wider spaced dots for the lower
    branch of the HFP calculation are unphysical; they lead to the
    corresponding wider spaced dots in Fig.~\ref{fig:phase-eos}.}
  \label{fig:eos}
\end{figure}

\begin{figure}[htpb]
  \centering
  \includegraphics[width=0.95\linewidth]{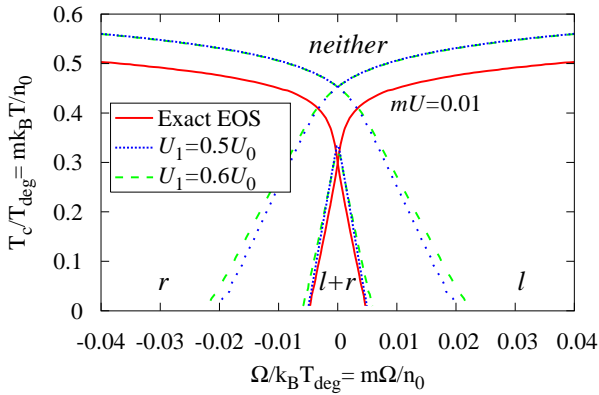}%
  \caption{(Color online) Critical temperature vs magnetic field (both
    measured in units of $k_B T_{\text{deg}}=n_0 / m$), at fixed total
    density.  Labels $l$, $r$, $l+r$ and \textit{neither} state which
    polarizations are superfluid.  Red solid line: Monte Carlo results
    valid when $U_1=0.5 U_0$. Blue dotted: HFP results for the same
    parameters.  Green dashed: HFP for a weak inter-polarization
    interaction.  The HFP lines at intermediate temperatures, with
    wider spaced dots and dashes, are unphysical.}
  \label{fig:phase-eos}
\end{figure}

The equation of state for a single component Bose gas, calculated
using a Monte Carlo (MC) method, are described in
Ref.~\onlinecite{prokofev02}.
In terms of the single component equation of state, the density
for the two component model may be written:
\begin{math}
  \label{eq:full-eos-1}
  \rho_{\text{total}}(\mu,\Omega,T) = \rho_{1}(\mu-\Omega,T) +
  \rho_{1}(\mu+\Omega,T),
\end{math}
and so the critical temperatures at field $\Omega$ are the solutions
of the equation:
\begin{equation}
  \label{eq:full-eos-2}
  \rho_{\text{total}} = \rho_1(\mu_{c1}, T_c) + \rho_1(\mu_{c1} \pm 2 \Omega, T_c).
\end{equation}
In two dimensions, the equation of state can be written $\rho_1(\mu,
T) = T \tilde{\rho}_1(\mu/T)$ and the critical chemical potential
scales as $\mu_{c1} = x_{c} T$.
Given $\tilde{\rho}_1(x)$ (as shown in Fig.~\ref{fig:eos}) and $x_c$, it is
straightforward to find $T_c(\Omega)$ for the two component case:
The two-component equation, Eq.~(\ref{eq:full-eos-2}), becomes 
\begin{equation}
  \label{eq:two-comp-eqn}
  \rho_{\text{total}} = 
  T_c\left[
    \tilde{\rho}_1(x_c) + \tilde{\rho}_1(x_c \pm 2 \Omega/T_c)
  \right] \equiv T_c F(\Omega/T_c).
\end{equation}
The phase boundary calculated from this equation of state is shown by
the solid line in Fig.~\ref{fig:phase-eos}.
The solution as $T_c\to 0$ requires $\tilde{\rho}_1 \to \infty$, which
for the MC calculation occurs at a finite $\Omega$ as $\Omega/T_c \to
\infty$.

Physically, the phases marked $l$ and $r$ correspond to pure circular
polarizations, while the phase $l+r$ will have an elliptical
polarization in general, and at $\Omega=0$, where the densities of $l$
and $r$ match, it will be linear.
Ref.~\onlinecite{rubo06} showed that at $T=0$, at the critical $\Omega$
separating $l$ from $l+r$, there is a quadratic gapless mode
corresponding to fluctuations of $r$.
It was suggested there that this mode prevents superfluidity, and so
suppresses the transition temperature to zero at this point.
However, this mode corresponds to excitations of $r$ which have
decoupled from those of $l$ at this point, and so a force acting only
on $l$ can still have a superfluid response; this is obviously the
case when $U_0=2U_1$, but remains true even for non-zero interactions.

\subsection{General case, $U_0 \ne 2 U_1$}
\label{sec:general-case-u_0}

For $U_0 \ne 2 U_1$, one would either need to perform new numerical
simulations, or find an appropriate perturbation scheme.
The simplest scheme one might consider is the Bogoliubov
approach,  which considers only the self energy due to
interactions with the condensate.
This approach is however incapable of describing transitions
such as that between $l+r$ both condensed and just $l$ condensed.
In the phase with just $l$ condensed, as $\mu \to \Omega$, the density
of $r$ will diverge, and so the density of $l$ (and hence the critical
temperature) will go to zero.
Considering this Bogoliubov approximation as applied to the case of
$U_0=2U_1$, this zero of the critical temperature can be understood as
follows:
If one plots the Bogoliubov approximation to the equation of state
$\tilde{\rho}_1(x)$, there is an unphysical divergence as $x=\mu/T \to
0^-$, while the exact equation of state is smooth.
It is clear that an unphysical divergence of $\tilde{\rho}_1(x)$ at a
finite value of $x$ leads to an unphysical extra solution of
Eq.~(\ref{eq:two-comp-eqn}) at $\Omega=0,T=0$.
This would produce a phase diagram like that of Ref.~\onlinecite{rubo06},
but is an artefact of neglecting the self energy due to the
non-condensed particles.

A better approach is the Hartree-Fock-Popov (HFP) method, see e.g.
Refs.~\onlinecite{kagan00,svistunov90}, which includes self energies due to
the population of thermal and quantum fluctuations, but neglects
anomalous correlations $\langle l^{\dagger} l^{\dagger}\rangle$ etc.
By including self energies due to fluctuation populations, there is no
divergence of density, as a large density $\rho^r$ would lead to a
large self energy, reducing the effective chemical chemical potential
for $r$.

The HFP method divides each polarization into a quasi-condensate
density, $\rho_0^{l,r}$ and fluctuation density $\rho_f^{l,r}$.
Both densities contribute to the gap equation:
\begin{align}
  \label{eq:gap1}
  U_0 (\rho_0^l + 2 \rho_f^l) + (U_0 - 2U_1) (\rho_0^r + \rho_f^r) &= \mu + \Omega \\
  \label{eq:gap2}
  U_0 (\rho_0^r + 2 \rho_f^r) + (U_0 - 2U_1) (\rho_0^l + \rho_f^l) &= \mu - \Omega
\end{align}
As in Ref.~\onlinecite{svistunov90}, the fluctuation density is found
from the correlation function $\left< l^{\dagger}(\vect{x}) l^{}(0)\right> =
\tilde{\rho}^l(\vect{x}) \exp[-\Lambda^l(\vect{x})]$.
The exponent $\Lambda^l(\vect{x})$ describes the power law decay of
correlations at long distances, and the prefactor
$\tilde{\rho}^l(\vect{x})$ describes decay from $\rho_0^l + \rho_f^l$
at $\vect{x}=0$ to the quasi-condensate density $\rho_0^l$ at
intermediate distances.
Thus $\rho_f^l = \tilde{\rho}^l(0) - \tilde{\rho}^l(\infty)$.
The correlation function is found from the the effective action for
density and phase fluctuations.
Writing $l=\sqrt{\rho_0^l + \pi^l}e^{i\theta^l}$ (and similarly for
$r$), one may define $\delta\Psi^{\dagger}=( \theta^l, \pi^l, \theta^r,
\pi^r)$, in terms of which:
\begin{align}
  \delta S 
  &= 
  \sum_{\omega,k} 
  \delta\Psi^{\dagger}(\omega,k) \mathcal{G}^{-1}(\omega,k) \delta\Psi(\omega,k)
  \nonumber\\
  \label{eq:inverse-full-greens}
  \mathcal{G}^{-1}
  &=
  \left(
    \begin{array}{cccc}
      2 \rho_0^l \epsilon_k & - \omega 
      & 
      0 & 0 
      \\
      \omega & U_0 + \frac{\epsilon_k}{2 \rho_0^l}
      &
      0 & (U_0 - 2U_1)
      \\
      0 & 0 
      &
      2 \rho_0^r \epsilon_k & - \omega 
      \\
      0 & (U_0 - 2U_1)
      &
      \omega & U_0 +  \frac{\epsilon_k}{2 \rho_0^r}
    \end{array}
  \right).
\end{align}
Because of the spinor structure, the current-current response
functions (and hence the superfluid density) have a tensor structure.
For example, left superfluid density is given by 
$\rho_s^l = (\rho_0^l + \rho_f^l) - \rho_n^l$
 where $\rho_n^l = m \chi_{T}^{ll}$ is the
normal density, found from the transverse part of the current-current
response function:
\begin{equation}
  \label{eq:define-current}
  \chi^{\alpha\beta}_{ij} 
  =
  \sum_{\omega,k} \text{Tr} \left[
    \mathcal{G}(k,\omega) \gamma^{\alpha}_i(k,k)
    \mathcal{G}(k,\omega) \gamma^{\beta}_j(k,k)
  \right],
\end{equation}
and $\gamma^{\alpha}_i$, the current vertex is (for $\alpha=l$):
\begin{equation}
  \label{eq:define-vertex}
  \gamma^l_i(k,k)
  =
  \frac{k_i}{m}  
  \left( 
    \begin{array}{cc}
      \sigma_2 & 0 \\ 0 & 0
    \end{array}
  \right),
  \qquad
  \sigma_2=
  \left(
    \begin{array}{rr}
      0 & -i \\ i & 0
    \end{array}
  \right)
\end{equation}

The above method gives the total density and superfluid densities
when both polarizations are condensed, and so gives the lower temperature
boundaries shown in Fig.~\ref{fig:phase-eos}.
The HFP method can also be adapted to find the higher temperature
boundary between a single polarization condensate and the normal
state.
Consider the case when only $l$ is condensed; in this case the total
density is $\rho_{\text{total}}=\rho_0^l + \rho_f^l+\rho_f^r$, and the
gap equation is just Eq.~(\ref{eq:gap1}) with $\rho_0^r=0$.
With a single condensate in the HFP approximation, there is no
coupling between the fluctuations of the $l$ and $r$ polarizations,
but only mean-field shifts from the quasi-condensate density, hence,
the inverse Green's function for fluctuations of $l$ is given by the
top left $2\times2$ block of Eq.~(\ref{eq:inverse-full-greens}).
For fluctuations of $r$, one has $\rho_f^r =\sum_k
n_B(\epsilon_k+\tilde\Sigma)$, where $\tilde\Sigma$ incorporates both
self energy and chemical potential, and is given by 
\begin{displaymath}
  \tilde\Sigma =
  2U_0 \rho_f^r + (U_0 - 2U_1) (\rho_0^l + \rho_f^l) - (\mu - \Omega).
\end{displaymath}
Eliminating $\mu$ using the gap equation, Eq.~(\ref{eq:gap1}), this
becomes:
\begin{displaymath}
  \tilde{\Sigma} = (U_0 + 2U_1) \rho_f^r - 2U_1 \rho_0^l - (U_0+
  2U_1) \rho_f^l + 2\Omega.
\end{displaymath}

Figure~\ref{fig:phase-eos} clearly shows that the HFP approximation
cannot accurately describe the entire phase boundary, even when
$U_0=2U_1$.
Although the low temperature boundary calculated by the HFP and exact
equation of state match well, the high temperature boundaries match
less well, and worse yet the critical temperatures do not join at
$\Omega=0$.
This failure is because the HFP equation of state is discontinuous at
the transition, and so is inaccurate when the high temperature
transition is close to the critical region of the low temperature
boundary.
However, far from the critical region the HFP approximation is
reasonable, i.e.\ at least as good as the HFP approximation would be
for a single component gas.

Despite its failings, the HFP method is valuable; firstly it shows
that the effect of weak inter-polarization interactions on the
critical temperature is small.
Secondly it reveals an intriguing property of the transverse
current-current response function at zero temperature.
In a single component condensate this vanishes, meaning the entire
system is superfluid.
For coupled polarizations it does not vanish, but instead one has
$\chi_{T}^{ll}=\chi_{T}^{rr} =-\chi_{T}^{lr}=-\chi_{T}^{rl} \ne 0$.
This identity ensures that a force acting on the total density has a
longitudinal (i.e.\ superfluid) response, but a force acting on only
one polarization need not.
This is expected, since there is overall Galilean invariance for a
change of velocity of both polarizations, but not under changes of the
velocity of just one polarization.
This zero-temperature transverse response is however small, $m
\chi_T^{ll} / \rho_{\text{total}} \propto m U_0$, as it arises due to
the quantum condensate depletion.

\section{Symmetry breaking effects in non-ideal cavities}
\label{sec:symm-break-effects}

The discussion so far is based on separate BKT transitions associated
with the proliferation of each kind of vortex; i.e.\ the effective
action is that of an $XY$ model:
\begin{equation}
  \label{eq:xy-model}
  \frac{S}{k_B T} = - \sum_{\left< ij \right>} \left[ 
    K^l \cos(\theta^l_i - \theta^l_j)
    +
    K^r \cos(\theta^r_i - \theta^r_j)
  \right].
\end{equation}
where $K^{l,r} = \rho_{s}^{l,r}/(m k_B T)$.
This section discusses some possible corrections to this effective
action that may change the critical behavior.
Three effects that are considered in detail:
The first is short range attraction between vortices of $\theta^l$ and
$\theta^r$ due to density-density interactions, which are already
present in the model of Eq.~(\ref{eq:hamil-fg}).
The other two effects concern reductions in symmetry present
in real cavities.
Even in an ideal quantum well, the symmetry group for Zinc-Blende
structures is not cylindrical, but $D_{2d}$\cite{ivchenko97}; this
means there are a preferred pair of axes, and so interactions between
polarizations of light do not have complete rotation
symmetry~\cite{rubo06}.
When there is asymmetry between the quantum well interfaces, this
symmetry is yet further reduced to
$C_{2v}$~\cite{aleiner92,winkler03:book}, leading to a preferred
linear polarization, causing a splitting of the quadratic polarization
terms.
Such a splitting can be induced by applying an electric field along
the growth direction ~\cite{malpuech06}; even without an applied
field, such a reduction of symmetry is observed in current
experiments~\cite{kasprzak06:nature,kasprzak07}.
These two types of symmetry reduction lead to perturbations that
prefer certain phase relationships between the $l$ and $r$ fields, and
so modify the order parameter space, hence change the behavior of the
phase transitions.
Another type of symmetry reduction concerns splitting between
transverse electric (TE) and transverse magnetic (TM) modes due both
to the cavity and exciton-photon
coupling~\cite{panzarini99,kavokin04,kavokin07}; this leads to an
interesting coupling between gradients of $l$ and $r$
fields\cite{rubo08}, which may shift the transition but does not
change the symmetry of the order parameter space so is not considered
in detail here.
The next section describes how the above considerations should be
incorporated as perturbations to Eq.~(\ref{eq:xy-model}), and
Sec.~\ref{sec:effect-pert-phase} then discusses their effect on the
phase boundary.

\subsection{Nature of perturbations to Eq.~(\ref{eq:xy-model})}
\label{sec:nature-pert-eq}

The short-range attraction between opposite vortices arises because
when $U_0 \ne 2 U_1$, a vortex of one polarization is associated with
a density modulation of the other (see Ref.~\onlinecite{rubo07}).
Thus a configuration with a vortex in each polarization has a lower
energy when these vortices are colocated, independent of whether the
phase windings of each polarization are aligned or anti-aligned.
This energy difference $\Delta E$ is finite, and numerical analysis
for $\Omega=0$ and small $(2U_1-U_0)/U_0$ gives $\Delta E/E_c \propto
(2U_1 - U_0)/U_0$, where $E_c$ is the core energy of a single vortex.
This effect can be represented in the Kosterlitz-Thouless scenario by
ascribing a larger fugacity to hybrid vortices (where both $l$ and $r$
wind) than to single vortices.

The reduction of symmetry in real crystals leads to terms that couple
the phase of $l$ and $r$ fields.
Breaking the symmetry to $D_{2d}$ means there are two orthogonal
preferred directions of polarization; since these orthogonal
directions are equivalent, this does not produce splitting in the
quadratic terms, but as discussed in Ref.~\onlinecite{rubo06}, there
is splitting due to interactions that may be written as $ |\psi_x|^4 +
|\psi_y|^4$; using Eq.~(\ref{eq:param}) this is proportional to
\begin{displaymath}
  |l+r|^4 + |l-r|^4 = 2 (|l|^2 + |r|^2)^2 + 4 (l^{\ast} r + l r^{\ast})^2.
\end{displaymath}
Breaking the symmetry yet further to $C_{2v}$ means favoring a
specific linear polarization, and this will introduce a splitting at
quadratic order; such a term would take the form $(l^{\ast} r e^{i
  \chi_0} + \text{H.c})$.
Both of the above terms can be written in the notation in
Eq.~(\ref{eq:xy-model}) as a term 
\begin{displaymath}
\delta H_p = \Delta_p \sum_i
\cos[p(\theta_i^l - \theta_i^r)],
\end{displaymath}
where $p=2$ describes the reduction to $D_{2d}$ and $p=1$ the
reduction to $C_{2v}$.

\subsection{Effect of perturbations on phase diagram}
\label{sec:effect-pert-phase}

The generalization of Eq.~(\ref{eq:xy-model}), including the
perturbation $\delta H_p$ was first discussed by~\citet{granato86}.
They included hybrid vortices, but only with aligned phase windings.
Their discussion is based on studying the renormalization group (RG)
flow in the Coulomb gas formulation of the model, where the dynamic
variables are the positions of vortices.
Using their formalism it is straightforward to show that when
$\Delta_p=0$, the effect of hybrid vortices is unimportant:
While in principle hybrid vortices can generate long-range
interactions between left and right vortices, this does not occur if
the energy of aligned and anti-aligned hybrid vortices are equal,
which is the case here~\footnote{TE-TM splitting can however
modify this; see Ref.\onlinecite{rubo08}.}.
This means that if $\Delta_p=0$ the transition is exactly the scenario
assumed above, where a BKT transition occurs if either species of
single vortex proliferates.

\begin{figure}
  \centering
  \includegraphics[width=0.9\columnwidth]{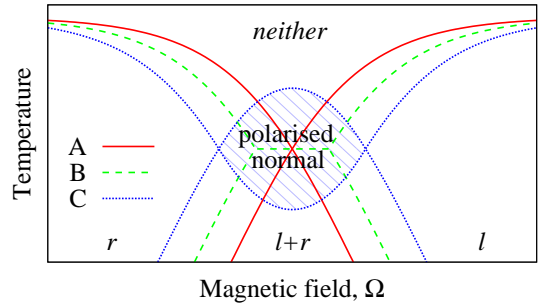}
  \caption{(Color online) Possible topologies of phase boundary in the
    presence of $p=2$ (i.e.\ quantum well) symmetry breaking; A and B
    reflect Ref.~\onlinecite{granato86}, C contains the possible
    polarized but non-superfluid phase suggested by
    Ref.~\onlinecite{hasenbusch05}.}
  \label{fig:topology}
\end{figure}

As the hybrid vortices alone have no significant effect, the model is
exactly that discussed by~\citet{granato86}.
Unfortunately perturbative RG cannot adequately describe the model
with $\delta H_p$ for $p\le 4$; the RG calculation
always tends to a phase in which some species of vortex or dual vortex
proliferate, and perturbation theory breaks down~\cite{nienhuis87}.
In addition, since one possible scenario discussed below involves two
closely separated phase transitions, numerical approaches are
challenging, as large system sizes are needed to prevent diverging
correlation lengths near one phase boundary masking effects of the
other~\cite{hasenbusch05}.
For this reason, rather than a definite conclusion, the remainder of
this article discusses the various possible scenarios
(Fig.~\ref{fig:topology}) that have been proposed for $\delta H_{2}$,
and their consequence for the polarized condensate.
\citet{granato86} suggest two possible topologies of the phase
diagram: either the same as without $\delta H_2$ (option A), or with a
region for $K^l\simeq K^r$ where there is a direct transition from the
$f+g$ condensate to the uncondensed state (option B).
Their suggestion was based on an argument~\cite{yosefin85} that when
$K^l=K^r$ there can be only one transition.
However, numerical simulations~\cite{hasenbusch05} of a closely
related model~\cite{yosefin85} suggest that at $K^l=K^r$ there are two
close but separate transitions: A higher temperature Ising transition
where $\theta^l-\theta^r$ becomes locked at either $0$ or $\pi$, and a
lower temperature BKT transition where power law correlations of
average phase occur.
For polaritons, this scenario (option C) implies an intermediate phase
with linear polarization but no superfluidity.
While polariton experiments might discriminate between these
scenarios, this would be made difficult by the close spacing between
the transitions, and the need for large homogeneous systems to avoid
finite size effects.

\section{Conclusions}
\label{sec:conclusions}

The phase diagram of polarized polariton condensate contains
transitions between regions of circularly polarized condensates and
elliptical polarizations.
Comparing numerical and perturbative methods when the two
polarizations decouple reveal limitations for the perturbative methods
near the critical point, but they suggest that weak interactions
between opposite polarizations have a small effect on the phase
boundary.
Terms which break the symmetry between different linear polarizations
can change the topology of the phase diagram; such effects may however
be very small, but present the possibility of using spinor condensates
for experimental investigation of the topology of this phase boundary.

\begin{acknowledgments}
  I would like to thank N.~R.~Cooper, P.~R.~Eastham, P.~B.~Littlewood
  for useful discussions, B.~Svistunov for providing
  Ref.~\onlinecite{svistunov90}, Y.~Rubo and A.~Kavokin for comments
  on an earlier draft, and Pembroke College, Cambridge for financial
  support.
\end{acknowledgments}

\end{document}